\documentclass[12pt]{article}
\usepackage{amssymb}
\usepackage{epsfig}
\usepackage{graphicx}
\usepackage{amsmath}
\usepackage{verbatim}
\title{Black Holes in the Presence of Cosmological Constant and Large $N$ Brane World}

  \author{{\small Mingxing Luo \footnote{Email adress:
  luo@zimp.zju.edu.cn} and Sibo Zheng\footnote{Email adress:
  zhengsibo@zimp.zju.edu.cn}  } \\
 {\small \it  Zhejiang Institute of Modern Physics, Department of Physics,}\\
  {\small \it Zhejiang University, Hangzhou 310027, P.R. China}\\
  }

\date{}
\begin{document}
 \maketitle

\abstract{Analytic form has been obtained for four-dimensional black holes with a minimal Hawking temperature
in a theory with cosmological constant, dilaton and gauge fields.
In general dimensions, black hole solutions are shown to exist and their asymptotic behaviors are obtained.
In theories of ten dimension,
$N$ coincident D3-branes as the boundary of an $AdS_5$ space are constructed by embedding black D3-branes,
with a five-dimensional compactified space of negligible size if $N$ is large,
which provide natural realizations of the Randall-Sundrum scenario.
For this $AdS_{5}$ background, the cosmological constant is a higher order perturbation
and its effect on the spectra of standard model fields on the branes can be calculated.}
\newpage

\section{Introduction}

Black hole solutions have been found in gravity \cite{ Gibbons:1988,Gibbons:1995,Maeda:1988}
and in low energy effective theories of Type IIB strings  \cite{andy:1991,andy:1992}.
Geometrically, all of them correspond to (anti-)de Sitter space.
Important applications of these black hole solutions are realized in $AdS_{5} \times S^{5}$,
in which there are $\mathcal{N}=4$ conformal field theories ($CFT_{4}$)
living on their boundaries by the principle of holography \cite{susskind,Maldacena:1997,Polchinski:1995,witten:19952}.
Large $N$ limit and weak coupling gravity is one way to guarantee the stability of this correspondence.
From which, quantities of strong coupling Yang-Mills gauge theories can be calculated by
classical approximations of gravitational theories in the bulk with appropriate boundary conditions
\cite{witten:1998,Polyakov:1998}.

In this paper, we will analyze gravitational theories
with gauge fields, dilatons, and a negative cosmological constant $\Lambda$.
A negative $\Lambda$ is consistent with supergravity \cite{Townsend:1977},
so some of these theories can be embedded into low-energy effective supergravity theories.
Black hole solutions found in Einstein-Maxwell model with cosmological term
\cite{Maeda:1988,Santos:0412,Tanay:0406} and black membrane
solutions including $F_{2}$ and non-zero dilaton field suggest
that we can find similar black hole solutions in these systems.

Indeed, we will find such black holes and determine their asymptotic behaviors.
They have gauge charges of a minimum critical value, or a minimum mass if it is a BPS state,
which is consistent with black hole thermodynamics and entropy theories \cite{hawking:1983,Harada:2006}.
Furthermore, we can construct compactified $N$ coincident D3-branes
from ten-dimensional gravitational theories,
in which dilaton potential provides an effective cosmological constant on the brane world \cite{Grojean:9910}.
The stability of the brane world can be kept at large $N$,
in which Hawking radiation in $AdS$ and loop corrections of gauge fields are both of order $1/N$.
The asymptotic $ AdS_{5}$ geometry arises naturally in these models.
The Kaluza-Klein compactified space can be tuned to the Planck scale by adjusting $N$.
We can find that these systems provide natural realizations of the Randall-Sundrum (RS) scenario
\cite{Arkani-Hamed:19981,Arkani-Hamed:19982,Randall:19981,Randall:19982}.

In section 2, black hole solutions will be obtained in terms of a set of parameters by making an explicit ansatz.
The positivity of the gauge charge $Q$ constrains the values of these parameters,
and $Q$ itself has a minimally critical value.
To manifest the $AdS_{5}$ geometry,
in section 3 we redefine fields as in \cite{andy:1991} to restructure the ten-dimensional theories,
and make necessary coordinates transformations near the horizon $r_{0}$ to extract the RS metric.
The mass spectra of four-dimensional scalar and $U(1)$ gauge field living on the branes are calculated
via perturbation theory.
These results are consistent with those from the analysis of gauged supergravity
\cite{Freedman:0009,Brandhuber:0010,Freedman:199904},
though they come from very different settings.

\section{Black holes in the presence of cosmological constant}
Black hole solutions in gravity and string theories were found long ago
\cite{Gibbons:1988,Gibbons:1995,Maeda:1988,andy:1991,andy:1992}
and extended to black p-brane solutions.
One may construct an action of D-dimensional $(4\leq D< 10)$ low-energy effective supergravity,
totally composed of bose modes.
In models containing a two-form Maxwell field $F_{2}$ and dilaton field $\phi$,
a cosmological constant can be included \cite{Gibbons:1988}.
A negative cosmological constant can be consistent with supersymmetry\cite{Townsend:1977} in some dimensions.\footnote
{One gets a fully supersymmetric theory if appropriate fermionic modes are added to the action.}
Black hole solutions in these systems were studied numerically in four dimension \cite{Poletti:1994}.
In this paper, we will analyze such solutions in arbitrary dimensions with an emphasis on their asymptotic behavior.
The low-energy effective action is
\begin{eqnarray}
\label{simpleaction}
 S_{eff}=\int d^{D}x
 \sqrt{-\tilde{g}}\left[\left(\tilde{R}-2\Lambda-\frac{1}{2}(\triangledown\tilde{\phi})^2\right)
  -\frac{2e^{\beta\tilde{\phi}}}{(D-2)!}F^{2}\right]
\end{eqnarray}
here the $(D-2)$-form gauge fields $F$ satisfies $dF=0$,
the tilde indicates that we are in the Einstein frame,
and the factor $2$ in front of the cosmological constant $\Lambda$ is chosen for convenience.
The equations of motion follows immediately
\begin{eqnarray}
\label{monopole}
\triangledown^{\mu_{1}}(e^{\beta\tilde{\phi}}F_{\mu_{1}\ldots\mu_{D-2}})=0\\
\label{dilaton}
 \triangledown^{2}\tilde{\phi}=\frac{2\beta}{(D-2)!}e^{\beta\tilde{\phi}}F^{2}\\
\label{gravity}
  \tilde{R}_{\mu\nu}=\Lambda
  \tilde{g}_{\mu\nu}+\frac{1}{2}\triangledown_{\mu}\tilde{\phi}\triangledown_{\nu}\tilde{\phi}&+&
   \frac{2}{(D-3)!}e^{\beta\tilde{\phi}}F_{\mu\lambda_{1}\ldots\lambda_{D-3}}F_{\nu}^{\lambda_{1}\ldots\lambda_{D-3}}\nonumber\\
   &-&\tilde{g}_{\mu\nu}\frac{2(D-3)}{(D-2)(D-2)!}e^{\beta\tilde{\phi}}F^{2}
\end{eqnarray}
where $\beta$ can be viewed as the coupling constant between the dilaton field and $F$ fields.
Without $\Lambda$, Eqs. \eqref{monopole}, \eqref{dilaton}, \eqref{gravity}
have a set of consistent gauge charged black hole solutions\cite{andy:1991,andy:1992}.
We will now find analogs with $\Lambda$.
To obtain spherically symmetric, static solutions, we make the following ansatz for the metric
\footnote{We have chosen our coordinate such that
$d\Omega_{D-2}=d\theta_{1}^{2}+sin\theta_{1}^{2}d\theta_{2}^{2}+\cdots+sin\theta_{i-1}^{2}d\theta_{i-2}^{2},
(i=2,\cdots,D)$.}
\begin{eqnarray}
\label{metric}
 d\tilde{s}^{2}_{D}=-\lambda dt^{2}+\frac{dr^{2}}{\lambda}+Rd\Omega_{D-2}
\end{eqnarray}
where $\lambda, R$ are functions of only $r$.
More specifically, we consider only monopole solutions with gauge charge $Q$:
\begin{equation}
 F=Q\epsilon_{D-2}, \ \ \ \ \
Q=\int_{S^{D-2}}F,
\end{equation}
where $\epsilon_{D-2}$ is the unit volume element of $S^{D-2}$.
One sees easily that Eq. \eqref{monopole} is automatically satisfied, independent of $\phi$ and
\begin{eqnarray}\label{strength}
F^{2}=F_{\lambda_{1}\cdots\lambda_{D-2}}F^{\lambda_{1}\cdots\lambda_{D-2}}=\frac{(D-2)!Q^{2}}{R^{D-2}}.
\end{eqnarray}

As there are two independent functions in Eq. \eqref{metric},
there should be only two independent equations in Eq. \eqref{gravity}.
For simplicity, we can pick up components $R_{00}$ and $R_{22}$, which can be written as
\begin{eqnarray}\label{components of metric}
R_{00}=\frac{\lambda(R^{\frac{D}{2}-1}\lambda')'}{2R^{\frac{D}{2}-1}},~~~~~~~~~
R_{22}=\frac{(D-3)}{R}-\frac{(R^{\frac{D}{2}-2}R'\lambda)'}{2R^{\frac{D}{2}-2}}
\end{eqnarray}
They yield
\begin{eqnarray}
\label{differantial equation1}
  \left[(D-3)\left(R^{\frac{D}{2}-2}R'\lambda\right)+R^{\frac{D}{2}-1}\lambda'\right]'=2R^{\frac{D}{2}-2}\left[(D-3)^{2}-(D-2)\Lambda
  R\right]
\end{eqnarray}
Furthermore, Eq. \eqref{dilaton} and $R_{00}$ imply,
\begin{eqnarray}
\label{differantial equation2}
  \left[R^{\frac{D}{2}-1}\left(\frac{2(D-3)}{\beta(D-2)}\tilde{\phi}^{'}\lambda+\lambda'\right)\right]'=-2\Lambda R^{\frac{D}{2}-1}
\end{eqnarray}
In addition to Eqs \eqref{differantial equation1}, \eqref{differantial equation2},
there is another constraint on the gauge charge $Q$ from Eq. \eqref{dilaton}
\begin{eqnarray}\label{chargeconstraint}
 2\beta
Q^{2}e^{\beta\tilde{\phi}}=R^{\frac{D}{2}-1}\left(R^{\frac{D}{2}-1}\lambda\tilde{\phi}'\right)'
\end{eqnarray}

\subsection{Decoupling the dialton field in $D=4$}

Our solutions will be greatly simplified if the dilaton field is a constant.
In this case, $\tilde{\phi}$ is decoupled from $F$,
since Eq. \eqref{chargeconstraint} imposes $\beta=0$
if we want to keep the monopole ansatz and $Q \neq 0$.
If $D=4$, we can actually get a simple analytic form of the metric
\begin{eqnarray}
\label{metricinfour}
\lambda(r)&=&\frac{1}{1+\Lambda}\left(1+\frac{c_{1}}{r}+\frac{c_{2}}{r^{2}}\right)-\frac{\Lambda}{3}r^{2}\nonumber\\
R(r)&=&(1+\Lambda)r^{2}\nonumber\\
Q^{2}&=&c_{2}(1+\Lambda)e^{2\phi}
\end{eqnarray}
where $c_{1}$ and $c_{2}$ are integral constants.
The metric Eq. \eqref{metricinfour} seems to have an extra singular point at $r\rightarrow\infty$.
But $R=g^{\mu\nu}R_{\mu\nu}$ and $R_{\mu\nu\rho\sigma}R^{\mu\nu\rho\sigma}$ are well behaved at $r\rightarrow\infty$,
so this singularity must be an artifact of our coordinate system.
$c_{2}$ is related to masses M of black holes.
If $c_{1}=0$, we get the charged Schwarzschild-AdS black hole.
Its horizon radius $r_{H}$ is
\begin{eqnarray}\label{horizon1}
 r_{H}=-\frac{\xi^{2}}{2}+\frac{\xi\sqrt{\xi^{2}-16c_{2}}}{2},~~~~~~~\xi^{2}=\frac{-3}{\Lambda(1+\Lambda)}
\end{eqnarray}
which is the only horizon and the Hawking temperature $T_{H}$ is
\begin{eqnarray}
\label{temperature1}
 T_{H}=\frac{1}{2\pi r_{H}}+\frac{r_{H}}{\pi\xi^{2}}
\end{eqnarray}
$T_{H}$ in Eq. \eqref{temperature1} have a minimum, i.e, $T_{H}\geq\sqrt{\frac{-2\Lambda}{3\pi}}$.
This is consistent with thermodynamical equilibrium theories of black holes in the AdS space \cite{hawking:1983}.
In the case $c_{1}\neq0$, we can have generalized charged Schwarzschild-AdS black holes,
which have at most four regular horizons.
By appropriately adjusting the value of $\phi$ and $c_{1}$, one can make $M=Q$ to obtain BPS states.

\subsection{Asymptotic black hole solutions in general dimensions}

$\lambda$ and $R$ behave differently with and without the cosmological constant term.
The asymptotic behavior at large radius will be modified greatly.
With the cosmological constant, as $r\rightarrow\infty$
\begin{eqnarray}\label{large r}
 \lambda \thicksim Cr^{2},~~~~~~R\thicksim r^{2},~~~~~~~~
 e^{\beta\tilde{\phi}}\thicksim r^{l}
 \end{eqnarray}
where $C=-\Lambda/(D-1)$, and the dilation field falls off quickly via a power law of the order $l=-2(D-2)$.
Consistent with the characteristics of black holes,
$\lambda$ and $R$ should be a set of functions that have singularity near the horizon $r_{0}$,
while conform to the asymptotic behavior Eq. (\ref{large r}) at large $r$.
The most general forms of asymptotic solutions
satisfying these conditions are ,
\begin{eqnarray}\label{metric in D-dimension}
\lambda(r)&=&H(r)^{x}+C(r-r_{0})^{2},~~~~~R(r)=r^{2}H(r)^{y},~~~~e^{\beta\tilde{\phi}}=F(r)^{z}\nonumber\\
H(r)&=&1-(\frac{r_{0}}{r})^{D-3},~~~~~~~~~~~~~~~F(r)=1-(\frac{r_{0}}{r})^{-l}
\end{eqnarray}
Obviously $H(\infty)=F(\infty)=1$.
The parameters $x,y,z$ will be determined by near horizon solutions
of Eq. \eqref{differantial equation1} to Eq. \eqref{chargeconstraint}.
There are two terms in $\lambda$, their relative importance around $\tilde{r}=r-r_{0} \sim 0$ depends on the value of $x$.
We first take the case $x<2$,  in which the cosmological constant is a perturbation near the horizon.
Plugging Eqs. (\ref{metric in D-dimension})
into Eqs. \eqref{differantial equation2}, \eqref{chargeconstraint}, we get
\begin{eqnarray}\label{parameter}
(D-2)\beta^2x=2z(D-3),~~~~~~~z=x+(D-2)y-2
\end{eqnarray}
and,
\begin{eqnarray}\label{charge1}
Q^{2}=\frac{z(D-3)^{z+2}r_{0}^{2(D-3)}}{2^{z+1}\beta^{2}(D-2)^{z}}
\end{eqnarray}
which is determined by the horizon $r_{0}$ and coupling constant.
The positivity of $Q$ implies that $z>0$. Furthermore, one gets $x>0$  from Eq \eqref{parameter}.

Now we consider Eq\eqref{differantial equation1}, which is quite involved.
There are four terms which can be potentially divergent when $\tilde{r}\rightarrow 0$.
To keep this in check, we need $y(\frac{D}{2}-1)+x-2$ to be positive.
That is,
\begin{eqnarray}\label{positivity}
 x+z>2
\end{eqnarray}
Under this condition, Eq\eqref{parameter} will be consistent asymptotic relations.
On the other hand, Eq.\eqref{parameter} and \eqref{positivity} imply that
there is a manifest minimally critical value $z_{0}=2\beta^{2}(D-2)/[\beta^{2}(D-2)+2(D-3)]$.
Since gauge charge $Q$ is a monotonic function of z, it thus has a minimum charge $Q(z_{0})$, i.e,
$Q(z)\geq Q(z_{0})$.
This constraint has not been observed in models without cosmological constant,
but agrees with recent works for BPS states by analyzing thermodynamics of critical black holes\cite{Harada:2006}.
This conclusion is valid for $x=2$ too,
in which case $Q$ contains a contribution from cosmological constant,
\begin{eqnarray}\label{charge2}
Q^{2}(z)=\left(\frac{D-3}{D-2}\right)^{z}\frac{z\left(C+\frac{D-3}{r_{0}}\right)^{2}r_{0}^{2(D-2)}}{2^{z+1}\beta^{2}}
\end{eqnarray}
If we ignore the cosmological constant, we get back to the original Eq. \eqref{charge1}.

If $x>2$, the cosmological term $Cr^{2}$ dominates near the horizon.
We will have a black hole, with its charge linear with $\Lambda$,
\begin{eqnarray}
Q^{2}=\frac{Cz(D-3)^{2y}r_{0}^{2(D-2)}}{2^{z+1}\beta^{2}(D-2)^{z}}
\end{eqnarray}
This case will not have an asymptotic AdS geometry and be of no interest to us.
From now on, we will concentrate on the case of $x<2$.
The metric Eq\eqref{metric in D-dimension} can be embedded into black p-brane solutions (see the next section),
$x,z$ and the coupling constant
will be determined by the embedding.
There is not much freedom as one had in \cite{andy:1991},
where one could write the ten-dimensional metric in a general form for Dp-brane\cite{Maldacena:1999},
\begin{eqnarray}\label{parameters}\nonumber
ds^{2}_{10}=\sqrt{H(r)}(-dt^{2}+dx_{i}^{2})+\frac{1}{\sqrt{H(r})}dr^{2}+Rd\Omega_{5},~~~(i=1,\ldots,p)
\end{eqnarray}
In \cite{andy:1992} this form is independent of dimension after we take $\alpha=0$.
In our case, we have used two functions $H(r)$ and $F(r)$ instead of one.
Different dimension will have different coupling constant and parameters $x,y,z$.

In summary, we have analytically obtained $D=4$ black hole solutions with constant dilaton field,
in which there exists a minimal Hawking temperature.
In general dimensions with non-decoupled dilaton field,
we have obtained gauge charged black hole solutions with cosmological constant and have determined their asymptotic behavior.
The behavior of black hole is greatly modified such that there exists a minimum gauge charge.
These results are consistent with thermodynamical and entropy theories of black holes.
In the next section we will discuss how to obtain a black p-brane solution.
Since $D-2=p+2$ for gauge fields,
the total dimension is $D+p=10$ for special interesting D3-brane.
We will concentrate on a ten-dimensional theory
that can construct an effective brane world based on the results in this section.

\section{ Large $N$ Brane World and Phenomenology}

In the last section, we have obtained black hole solutions
in models of D-dimensional low-energy effective theory with cosmological constant.
A cosmological constant cannot appear directly in ten or eleven-dimensional supergravity.
But it can be part of higher dimensional action of gravity by dimensional reduction.
It can also be subset of ten-dimensional low-energy effective theory,
in which a dilaton potential $U$ provides the cosmological constant.
The background metric can be obtained by Kaluza-Klein compactification.
Here, we will compactify five dimensions to construct an $AdS_{5}$ background,
in which case the reduction relations given in \cite{andy:1991} are still valid.
A finite large $N$ scenario can realize the localization of $D3$ branes
with the exponential form $e^{N^{\frac{1}{4}}\varphi}$.
This scenario implies an effective RS metric \cite{Arkani-Hamed:19981,Arkani-Hamed:19982,Randall:19981,Randall:19982}
even the dilaton and gauge fields are present,
which has been discussed in the frame of low-energy effective superstring theory \cite{Grojean:9910}.

The finite large $N$ scenario also means our ten-dimensional
gravitational action is a weak coupling model, which gurantees
that the Hawking radiation in $AdS$ and loop corrections of gauge
fields are of the order $1/N$. So the classical ground background
are valid for the non-supersymmetric model. For this
$AdS_{5}$ background, the bulk cosmological constant is higher
order perturbation, so one can analyze the spectra of
four-dimensional standard model (SM) fields on D3-branes
perturbatively.

\subsection{Large $N$-brane world}
Eq.(\ref{simpleaction}) can be obtained from a ten-dimensional gravitational theory of action in the string framework
\begin{eqnarray}
\label{action2}
 S=\int d^{10}x
 \sqrt{-g}\left[{e^{-2b\phi}}\left(R+f(b)(\triangledown \phi)^2 \right)-\frac{2e^{2\alpha\phi}}{(D-2)!}F^{2}+ U\right]
\end{eqnarray}
via the transformation $\tilde{g}_{MN}(Einstein)=e^{-\frac{b}{2}\phi}g_{MN}(string)$ when one sets,
\begin{eqnarray}
\label{function f}
f(b)=\frac{9b^{2}-1}{2}
\end{eqnarray}
The dilaton potential is $U(\phi)=-\Lambda e^{b_{*}\phi}$ ($b_{*}$
to be specified later).
Since $U$ is forbidden in ten-dimensional supergravity,
Eq.(\ref{action2}) is not part of a low-energy effective supersymmetric action.
The classical approximative solution of Eq.(\ref{action2}) is only effective and stable for weak coupling system.

Indeed, black D3-brane solutions are such a realization. From
Eq.(\ref{metric in D-dimension}) we see $e^{-\phi(z)}\sim 0$ near
the horizon $(z\rightarrow \infty)$ and $e^{-\phi(z)}\sim 1$ as
$z\rightarrow 0$. The effective coupling will be weak
near the horizon where the $D3-$branes locate. Besides, black
D-brane solutions of appropriate large $N$ implies that the
curvature function $R(1/r_{0})\sim R(1/N)$ of the black p-brane
geometry in the bulk is close to zero. So quantum fluctuations of
gravity itself, e.g. the Hawking radiation, are very small.
In this sense, the classical description of Eq(\ref{action2}) using
black p-branes is appropriate, similar to the $AdS_{5}\times
S^{5}$ black p-brane solution \cite{andy:1991} and cosmological
Schwarzschlid-AdS black holes of classical gravitational theories
\cite{cvetic:0306}. In general, the embedded metric of black
D3-branes can be written as \footnote{A review on reduction of
ten-dimensional theories to lower dimension can be found in
\cite{andy:1991}, \cite{Petersen:1999}. We used Eq.(3) and Eq. (4)
in \cite{andy:1991}, with modification: $\beta
A=\tilde{\phi}(\alpha b-\frac{3}{5})$ and $\beta
B=\tilde{\phi}(\alpha b+1).$}
\begin{eqnarray}\label{ten-metric}
ds^{2}_{10}&=&e^{A}d\tilde{s}^{2}_{D}+e^{B}\eta_{ij}dx^{i}dx^{j},~~~~~~~~~~( 1\leq i\leq p)\nonumber\\
&=&F^{\Delta}\left[-\left(H^{x}+C(r-r_{0})^{2}\right)dt^{2}+\left(H^{x}+C(r-r_{0})^{2}\right)^{-1}dr^{2}+F^{x}dx_{i}^{2}\right]\nonumber\\
 &+&r^{2}F^{\Delta}H^{y}d\Omega_{5}\nonumber\\
b_{*}&=&\phi^{-1}\left(\frac{10-p}{2}A+\frac{p}{2}B\right)-\frac{5b}{2}
\end{eqnarray}
where $\Delta=z(\alpha b-\frac{3}{5})/\beta^{2}$. Eq.(\ref{parameter}) has been used in the second line of above equation.
We make the following coordinates transformation,
\begin{eqnarray}\label{coordinate transformation}
\rho^{4}=r^{4}-r_{0}^{4},~~~~~~~~~ \rho=\frac{h^{2}}{z},~~~~ with~~r_{0}^{4}=K(z,\alpha)N
\end{eqnarray}
where $K(z,\alpha)$ is a positive constant coefficient related to $z$ and coupling $\alpha$ in Eq.(\ref{charge2}) with $D=7$,
and $N$ is the quantum number of charge $Q$.
We then obtain an asymptotic $AdS_{5}$ metric when $z\rightarrow\infty$,
\footnote{We have dropped a higher-order term in time coordinate,
which violates Lorentz invariance slightly in the bulk but has no effects on the branes.}
\begin{eqnarray}\label{adsmetric}
ds^{2}_{5}=\left(\frac{r_{0}}{z}\right)^{2}\left[\eta_{\mu\nu}dx^{\mu}dx^{\nu}+\left(1+\frac{r_{0}^{5}C}{16z^{3}}\right)^{-1}dz^{2}\right]
\end{eqnarray}
when the coupling $\Delta=-3/4$ and parameter $x=5/4$ are specifically taken.
Thus $b_{*}<0$ for stability. And the Kaluza-Klein compactified metric is,
\begin{eqnarray}\label{compactified manifold}
r^{2}F^{\Delta}H^{y}d\Omega_{5}=\left(\frac{5}{2}\right)^{-\frac{3}{4}}r_{0}^{2+4\vartheta}z^{-4\vartheta}d\Omega_{5},~~~~~\vartheta=\Delta+y
\end{eqnarray}
Unlike the p-brane solutions without cosmological constant
\cite{andy:1991,andy:1992} in which the scale of Kaluza-Klein compactified manifold is $r_{0}$,
the sign of $\vartheta$ in Eq.(\ref{compactified manifold}) can be changed by adjusting the positive $b$.
Specifically if
\begin{eqnarray}\label{b value}
b^2<{\frac{1}{10}}\left(3+\frac{8\Delta}{x}\right)^2
\end{eqnarray}
one has $\vartheta>0$ and the important constraint required by Eq(\ref{positivity}) can be satisfied.
So in the vicinity of
$z\rightarrow\infty$, the compactified five-dimensional manifold
is of negligible size. And Eq(\ref{adsmetric}) is a standard
anti-de Sitter metric that can be written in the RS form,
\begin{eqnarray}\label{Rs metric}
ds^{2}_{5}&=&\frac{1}{t^{2}}\eta_{\mu\nu}dx^{\mu}dx^{\nu}+\left(\frac{r_{0}}{t}\right)^{2}dt^{2}\nonumber\\
&=&e^{-2r_{0}\varphi}\eta_{\mu\nu}dx^{\mu}dx^{\nu}+r_{0}^{4}d\varphi^{2},~~~~~t=\frac{z}{r_{0}}=e^{r_{0}\varphi}.
\end{eqnarray}

One reads from Eqs. (\ref{coordinate transformation}) and (\ref{Rs metric}) that,
for finite large $N$, $z=r_{0}e^{r_{0}\varphi}$ approaches infinity exponentially for decent values of $r_0$.
And the scale of Kaluza-Klein compactified space is suppressed,
while the scale of fifth extra dimension $\varphi$ can be very large.
We can choose a slice of $AdS_{5}$,
i.e, $\varphi\in[\epsilon_{0},\pi]$ with $\epsilon_{0}$ a small positive number.
This case is very similar to the alternative compactification discussed in \cite{Randall:19981,Randall:19982},
although we begin at the Kaluza-Klein compactification in a more complicated model.
In conclusion, a generalized RS action including more massless modes can still have a consistent description
about the hierarchy between Plank scale $M_{Pl}$ and electro-weak scale $m_{EW}$,
if we replace finite large N coincident D3-branes instead of a single D3-brane.
For example the five-dimensional Plank mass related to four-dimensional Plank mass as
\begin{eqnarray}\label{plank mass}
M_{Pl}^{2}\sim 2(\pi-\epsilon_{0})M^{3}r_{0}^{4}e^{-2r_{0}\epsilon_{0}}
\end{eqnarray}
and the five-dimensional electro-weak mass related to effective four-dimensional mass for branes located at $\varphi=\pi$,
\begin{eqnarray}\label{electro-weak mass}
m\sim e^{-2r_{0}\pi}m_{EW}
\end{eqnarray}

\subsection{Spectra of SM fields: scalar field }

As shown in Eq.(\ref{adsmetric}), cosmological constant is a low-energy perturbation to flat $AdS_{5}$ background.
One can obtain the mass spectra of different
kind of fields living on the D3-brane world at small internal $\varphi$
or large $z\in [z_{0},\infty]$ with $z_{0}=r_{0}e^{r_{0}\epsilon_{0}}$ by perturbation.
Firstly, we take a massless scalar field as an example.
There is no need to consider the compactified space,
similar to the simplicity of s-wave in calculating the correlation functions \cite{Polyakov:1998}.
The action for $\phi$ then becomes,
\begin{eqnarray}\label{action of Ads}
S&=&\frac{1}{4\kappa^2}\int d^{10}x
\sqrt{g}\left[g^{MN}\partial_M\phi\partial_N\phi\right]\nonumber\\
&=&\frac{1}{4\kappa^2}\int d^4x \int_{z_0}^\infty dz\frac{1}{z^3}
\left[(1+\frac{r_0^5C}{16z^3})(\partial_z\phi)^2+\eta^{\mu\nu}\partial_\mu \phi\partial_\nu \phi \right]
\end{eqnarray}
Minimizing the action, one gets the equation of motion,
\begin{eqnarray}\label{equation of motion}
\left(1+{\tilde C \over z^3} \right)\tilde\phi''-\left({3 \over z}+{6\tilde C \over z^4}\right) \tilde\phi'-k^2\tilde\phi=0
\end{eqnarray}
here $\tilde{C}=r_{0}^{5}C/16$ and we have made a Fourier transformation of $\phi$ into momentum space $
\phi(x)=\int d^{4}k\lambda_{k}\tilde{\phi}(z)e^{ikx}$.
The contribution of cosmological constant is of high-order.
So this equation can be solved by perturbation.
Writing $\tilde{\phi}=z^{\frac{3}{2}}g$, Eq. \eqref{equation of motion} gives,
\begin{eqnarray}\label{perturbation}
\left(1+\frac{\tilde{C}}{z^{3}}\right)g-\frac{3\tilde{C}}{z^{4}}g'-\left(\frac{15}{4z^{2}}+\frac{33\tilde{C}}{4z^{5}}\right)g=k^{2}g
\end{eqnarray}
Without contributions from the cosmological constant,
modified Bessel functions $g_{0i}=\sqrt{z}K_{2}(k_{i}z)/\sqrt{N_{i}}$ solve the equation,
which form a set of complete eigenfunctions
\begin{eqnarray}
\int_{h_{0}}^{\infty}dz g_{0i}g_{0j}=\int_{h_{0}}^{\infty}dz
zK_{2}(k_{i}z)K_{2}(k_{j}z)=\delta(k_{i}-k_{i})
\end{eqnarray}
with renormalized factor $ N_{i}=\int_{z_{0}}^{\infty}zK_{2}^{2}(k_{i}z)dz$.
where $k_{i}$ is the eigenvalue.
Make a perturbation expansion for the solution of eq\eqref{perturbation}
\begin{eqnarray}
g=g_{0}+\sum_{i}b_{i}g_{0i}
\end{eqnarray}
Notice the perturbation potential
\begin{eqnarray}\label{potential}
V=\tilde{C}\left(\frac{1}{z^{3}}\partial_{z}^{2}-\frac{3}{z^{4}}\partial_{z}-\frac{33}{4z^{5}}\right)
\end{eqnarray}
To the first order, we have
\begin{eqnarray}\label{solution}
\tilde{\phi}_{k}&=&z^{\frac{3}{2}}\left[g_{0k}+\sum_{n}\frac{V_{nk}}{E_{0k}-E_{0n}}g_{0n}\right]\nonumber\\
V_{nk}&=&<g_{0n}\mid V\mid g_{0k}>
\end{eqnarray}
So the spectrum of the scalar field is
\begin{eqnarray}\label{spectrum}
m^{2}_{S}=-<g_{0}\mid V\mid
g_{0}>=-\tilde{C}\int_{z_{0}}^{\infty}dz
g_{0}^{*}\left(\frac{g_{0}''}{z^{3}}-
\frac{3g_{0}'}{z^{4}}-\frac{33g_{0}}{4z^{5}}\right)=\frac{e^{-5r_{0}\epsilon_{0}}C}{16}
\end{eqnarray}
Here we have taken $k^{2}=0$ limit for the unperturbed scalar field and
used the asymptotic form of modified Bessel functions $K_{\nu}(x)\rightarrow\frac{\pi}{2\sqrt{x}}e^{-x},(x\rightarrow\infty)$.
The excited mode is totally due to the cosmological constant.

\subsection{Spectra of SM fields: vector field }

Now we turn to gauge or vector fields.
Roughly speaking, we can expect similar mass contribution from the C-term,
as transverse components of the gauge field has similar equations of motion as scalar field.
We consider a simple $A_{\mu}$ to manifest our assumption.
The Lagrangian of 1-form $A=A_{\mu}dx^{u}$  gauge field is
\begin{eqnarray}\label{gauge action}
S_{gauge}=-\frac{1}{4}\int d^{10}x\sqrt{-g}
F_{\mu\nu}F^{\mu\nu } + \int F_{2}\wedge F_{2}\wedge A
\end{eqnarray}
The first term will contribute to the energy-momentum tensor of gravity sysyem
and is generally valid to take this term as a perturbation to eq\eqref{action2}.
The Chern-Simons term will not affect the equation of motion,
but contributes to n-point $(n>2)$ correlation functions.
Minimizing the action, we obtain the equation of motion
\begin{eqnarray}\label{vector field}
\left[\left(1+\frac{\tilde{C}}{z^{3}}\right)\partial_{z}^{2}-\left(\frac{2}{z}+\frac{\tilde{C}}{2z^{4}}\right)\partial_{z}+\partial_{i}^{2}\right]A_{j}=0
\end{eqnarray}
for transverse components of the vector field: $\partial_{i}A^{i}=0~(i,j=1,~\ldots, 4)$.
Making the substitution $A=z\tilde{A}$, one gets from Eq. (\ref{vector field}),
\begin{eqnarray}\label{vector field 1}
\left[\left(1+\frac{\tilde{C}}{z^{3}}\right)\partial_{z}^{2}+\left(\frac{\tilde{3C}}{2z^{4}}\right)\partial_{z}
-\left(\frac{2}{z^{2}}+\frac{\tilde{C}}{2z^{5}}\right)+\partial_{i}^{2}\right]\tilde{A}_{j}=0
\end{eqnarray}
Without the C-term perturbation, the equation is solved by modified Bessel functions of the order $\nu=3/2$
and vector field has a conformal dimension of $\Delta=3$.
This agrees with general conclusions for vector fields \cite{witten:1998,Freedman:199904}.
Similar to scalar fields, the C-term implies a small mass to the gauge field,
\begin{eqnarray}\label{mass of vector field}
m^{2}_{A}=\frac{e^{-5r_{0}\epsilon_{0}}C}{192}
\end{eqnarray}

In fact, the situation is rather similar to the five-dimensional gauged supergravity
with various extended supersymmetries in anti-de sitter space
\cite{Behrndt:199907,Cvetic:1996,Freedman:0009,Brandhuber:0010},
where the components of vector field can be written as a Schroedinger equation with general form
\begin{eqnarray}
\partial_{z}(e^{A}\partial_{z}V_{i})+m^{2}e^{A}V_{i}=0\nonumber
\end{eqnarray}
This Schroedinger equation has a potential $V=\frac{1}{4}(A')^{2}+\frac{1}{2}A''$,
which provides a result similar to eq\eqref{mass of vector field}.
In the theory of five-dimensional gauged supergravity,
there is another term associated with the breaking of non-abelian gauge symmetries in the potential.
In our case, there is no breaking of gauge symmetries.
However the superpotential $V_{s}$ corresponding to the bulk potential $V$ will contains
a higher order term from cosmological constant $\Lambda$,
which carries $\mathcal{R}$ charge $Q_{\mathcal{R}}\neq2$, thus breaks the $U(1)_{\mathcal{R}}$ symmetry.

\section{Conclusion}
In this paper we have discussed gauge charged black hole solutions in D-dimensional low-energy effective supergravity
with negative cosmological constant.
We have not added complex terms of fermion modes that keep the supersymmetries to the effective action.
Analytic solution is obtained in four dimension with a minimum Hawking temperature,
which is consistent with the thermodynamics of Schwarzschild-AdS black holes.
For general dimensions, we have obtained asymptotic solutions of a minimum gauge charge,
which means a minimum mass for a BPS state, similar to critical black holes.
These black hole solutions may break part or all supersymmetries,
but our analysis should still be reliable since our models are valid for classical approximation under large $N$ scenario,
similar to the cosmological Schwarzschild-AdS black holes.

And a ten-dimensional action including massless dilaton and gauge fields can be solved
by compactification based on the results in section 2.
Indeed, the scale of compactified space is negligible at finite large $N$ scenario.
From which, we can obtain an effective D3-brane world in more generalized cosmological model
that includes massless dilaton and form fields.
These systems provide natural realizations of the RS scenario.
And the spectra of SM fields in this background are consistent
with phenomenological constraints of effective four-dimensional D3-brane.

\section*{Acknowledgement}
It is a pleasure to thank J. H. Huang and W. S. Xu for discussions.
This work is supported in part by the National Science Foundation of China (10425525).

\end{document}